\documentstyle[preprint,prl,aps,psfig]{revtex}   

\begin{document}                                                                
\tighten
\title{Application of the Kerman-Klein method to the solution of a spherical 
shell model for a deformed rare-earth nucleus} 

\author {Pavlos Protopapas and Abraham Klein}
\address{Department of Physics, University of Pennsylvania, Philadelphia,
Pennsylvania, 19104-6396}

\date{\today}

\maketitle

\begin{abstract}

Core-particle coupling models are made viable by assuming that core
properties such as matrix elements of multipole and pairing operators
and excitation spectra are known independently.  
From the completeness
relation, it is seen, however, that these quantities are themselves
algebraic functions of the calculated core-particle amplitudes.  
For the deformed
rare-earth nucleus $^{158}$Gd, we find that these sum rules are 
well-satisfied for the ground state band, implying that we have found
a self-consistent solution of the non-linear Kerman-Klein equations.

\end{abstract}
\bigskip


\narrowtext   

1.  {\it Introduction}.  The phenomenological shell model remains the 
bedrock of nuclear structure physics \cite{Lawson,Brown}.  
In its standard form, 
we accept the empirically established notions of closed shell or magic
nuclei and of single particle or single hole excitations with respect
to these special cores (below, for brevity, we speak only of particles).  
To study the properties of nuclei which
are removed from these stable cores by two or more nucleons, one adds
residual two-particle forces.  To understand low energy behavior of low
and medium mass nuclei
one restricts the allowed 
single-particle excitations and residual interactions to the valence shell.
The solution of the resulting matrix diagonalization problem, which is
straightforward in principle, has been achieved only up to mass number $N=48$
\cite{Caurier} because of the rapid growth of the dimensionality of the 
Hamiltonian matrices.  Beyond that there are several possibilities:
One can study somewhat heavier nuclei with Monte Carlo
calculations \cite{Alhassid} that utilize the entire valence shell-model space.
One can also reduce the dimensionality of the Hamiltonian matrices
to tractable sizes in several ways, either
by utilizing only the lowest irreducible representations of relevant
approximate symmetries \cite{Covello,Draayer} or by applying the 
variation-Al method to a trial space suggested by the deformed shell model
\cite{Hara,Schmid}.  Except for the first example cited \cite{Covello}, 
these latter methods allow one to break the bounds of the valence shell
restriction.

More than three decades ago, A.\ K.\ Kerman and one of the authors
proposed an alternative to the standard linear approach to 
the shell model.   Originally designed as a method for restoring
the broken symmetry of mean field solutions \cite{K1},
it soon became clear
that it was a general formulation of quantum mechanics \cite{K2} 
that could also be
used to study the shell-model problem.  It was argued that especially in 
cases of well-developed collective motion, one could replace the 
linear methods that use a large basis of states by a non-linear
method involving a tractable set of states.  Early attempts 
to apply this method to semi-magic nuclei \cite{K4,K5} 
were at best only modestly successful and were not followed up.

Our aim in this letter is to reactivate the original program by 
reporting a successful application to a deformed nucleus in the 
rare earth region, $^{158}$Gd.  For such a nucleus not only is it 
technically impossible to apply the valence shell model based on
spherical single-particle excitations, but the restriction to the valence
shell itself fails badly \cite{Lawson,Draayer,Hara,Schmid}.  
Starting from a spherical shell model expanded
to include all orbits bound in a realistic (Wood-Saxon) single-particle
potential and a standard Hamiltonian widely applied for heavy nuclei, we 
describe a fully microscopic derivation of some of the properties of the ground
state rotational band including energies, charge and mass quadrupole
matrix elements, and pairing matrix elements.  This work was carried out
using results obtained from a semi-microscopic description, described 
below and referred to as CPC, of the low energy
properties of {\it odd} deformed nuclei.  The results reported for
$^{158}$Gd are almost certainly not special to this nucleus, nor is the 
method necessarily confined to the ground-state band.

2. {\it Model and method}.     As Hamiltonian we choose the form 
\begin{eqnarray}
H &=& \sum_{\alpha i}h_{ai}a^{\dag}_{\alpha i}a_{\alpha i} \nonumber \\
&& -\frac{1}{2}\sum_{ij}\sum_{LM_L}\kappa_{ij;L}Q^{i\dag}_{LM_L}
Q^j_{LM_L}   \nonumber \\
&& -\frac{1}{2}\sum_i\sum_{LM_L}g_{i;L}\Delta^{i\dag}_{LM_L}
\Delta^i_{LM_L}.  \label{ham}
\end{eqnarray}
In the first term $a$ and $a^{\dag}$ are the spherical shell 
model annihilation and
creation operators, $\alpha$ labels the principal and angular momentum
quantum numbers of states in a spherical Wood-Saxon
potential with spin-orbit coupling, the subscript $a$ the same set minus
the magnetic quantum number, $i$ distinguishes neutron from proton, 
and $h_{ai}$ are the eigenvalues in the respective wells.  The second term
is a sum of products of mass multipole moments with strengths
$\kappa_{ij;L}$; in the following we shall
retain only the most important terms, those with $L=2$, though in a more 
refined treatment, we should include $L=4$ \cite{Pav1}.
The last term is a sum of pairing 
interactions with strengths $g_{i;L}$ , of which we include only the 
dominant monopole, though here
the $L=2$ term can also be considered to be well established \cite{Hara}.
The adequacy of the Hamiltonian (\ref{ham}) as a representation
for those properties of a more realistic interaction that lead to 
collective behavior 
has been carefully documented in a recent investigation \cite{Zucker}.

To explore the consequences of (\ref{ham}), we calculate the commutator
of $a$ and of $a^{\dag}$ with the Hamiltonian and take the matrix elements
between states $|J\mu\nu\rangle$ of a chosen odd nucleus of mass number $N$,
with angular momentum quantum numbers $J,\mu$ and all other labels indicated
by $\nu$ and the corresponding states of the relevant even neighbors,
$|IMn(N\pm 1>$.  Suppressing charge quantum numbers, we encounter in the 
single-particle terms the coefficients of fractional parentage (CFP)
\begin{eqnarray}
V_{J\mu\nu}(\alpha;IMn> &=& <J\mu\nu|a_{\alpha}|IMn(N+1)> \label{cfp1} \\
U_{J\mu\nu}(\alpha;IMn) &=& <J\mu\nu|a_{\bar{\alpha}}^{\dag}|
IMn(N-1)>. \label{cfp2}
\end{eqnarray}
For the evaluation of a typical interaction term, consider, in an abbreviated
notation,
\begin{eqnarray}
&& \langle J|aQ_2|I(N+1)\rangle \nonumber \\
&&=\sum_{I'} V_J(I')\langle I'(N+1|Q_2|I(N+1)\rangle, 
\label{comp1} 
\end{eqnarray}
which involves only the completeness relation.  By this means the non-linear
terms are expressed as sums of products of terms in which one factor, the 
CFP, depends on the odd nucleus, whereas the other depends on the properties
of the even cores.
With a corresponding treatment
of the pairing interaction, we obtain equations with 
characteristics and properties that we now describe.

In addition to the CFP, which
are coupled by the pairing interactions, there occur in these equations
matrix elements of the mass quadrupole moments for two different, 
neighboring, even nuclei and matrix elements of the pairing interaction between the two neighbors. 
Together we refer to these as the core matrix elements.   The structure
of these equations, which is given in all detail in \cite{Pav2} and will not
be reproduced here, bears a striking resemblance to those of the 
Hartree-Bogoliubov mean-field theory, but in contrast to the latter our
equations are formally exact, conserving both angular momentum and particle
number.

If we assume that the core matrix elements are known, the
resulting equations are linear and define an Hermitian eigenvalue problem
for the energies of the odd nucleus relative to the average ground 
ground state energy of the neighboring cores.  In this interpretation, the
chemical potential of the odd nucleus and the excitation spectra of the even
neighbors are added to the list of quantities assumed to be known.  
The solutions are mutually orthogonal, and the normalization will be considered
below.   The possibilities inherent
in such a generalized semi-microscopic theory, first noted in \cite{K6}, was
first developed and applied by D\"onau and Frauendorf \cite{DF1,DF2}.  Recently
further development and applications have been
carried out by the authors in a series of papers of which the latest are 
\cite{Pav1,Pav2}.

It is vital to recall the approximations that are made
in order to make even the linear scheme workable.  The most important can
be understood by examination of (\ref{comp1}).  If the starting state of the
core, $|In\rangle$ belongs to a given low-lying rotational band, 
then we know that 
intra-band transitions are by far the dominant ones, though a few neighboring
bands provide some residual strength, and these are included in the 
calculations in order to satisfy the sum rule. 
Once the choice of core bands has been made,
there remains a vital question associated with the space of the 
single odd nucleon.  For all the 
examples done, we find that results for the observables of interest
have essentially converged when
three major shells are included.  For the purposes of the new results reported
in this letter, we have nevertheless done calculations that include all 
bound orbitals.   The reason for this will be explained below.

3. {\it Self consistency: particle number}.  
We now take the next 
step and ask whether the solutions of the CPC calculations that we have 
described
are in fact self consistent.   Naturally we have chosen to examine first
the most favorable case of a 
strongly deformed nucleus in a region where the properties to be tested 
vary slowly and smoothly with particle number.
We start with what may at first sight appear to be a trivial example, 
the conservation of particles.
The operator for the total particle number, 
\begin{equation}
  \hat N =\sum_{\alpha,i} a^{\dag}_{\alpha i}  a_{\alpha i} \label{num1}
\end{equation}
can be separated into a sum of four terms
\begin{equation}
  \hat{N} = \hat{N}_{\rm{p},+} +\hat{N}_{\rm{p},-} 
  +\hat{N}_{\rm{n},+}+\hat{N}_{\rm{n},-},   \label{num2}
\end{equation}
where the subscripts distinguish charge of the nucleons and parity of the 
single-particle orbitals.  Because we include in the Hamiltonian (\ref{ham})
only multipoles of even parity, each of these quantities
is conserved.  The even nucleus chosen for study plays the role of the heavier
of the two cores in a calculation \cite{Pav3} carried out for $^{157}$Gd,
an axially deformed nucleus with states $|IMK(N+1)\rangle$, where $K$ is the 
angular momentum of the band head.  Further discussion will be confined to the
ground-state band with $K=0$ and this quantum number will be suppressed.
We thus need the four sets of eigenvalues
\begin{eqnarray}
  N_{i, \pm} &=& 
       \sum_{\alpha \pm}
       \langle IM(N+1) |  a^{\dag}_{\alpha i\pm} a_{\alpha i\pm}|
       IM(N+1) \rangle \nonumber \\
       &=& \sum_{\alpha \pm, J\mu\nu}
       V^{i~\star}_{J\mu\nu}(\alpha;IM) V^{i}_{J\mu\nu}(\alpha;IM),\label{num3}
\end{eqnarray}
each of which should be independent of $I$, as it automatically is 
independent of M.  To evaluate these sum rules, one needs the CFP for the
neutron levels of $^{157}$Gd of both parities, which had been obtained earlier
\cite{Pav3} and the proton levels of $^{157}$Eu, which were obtained for 
present use.
The results are shown below in Table~\ref{tab:Num}.  It has been verified that
the sums in (\ref{num3}) depend on $I$ only in the third decimal place.

We consider the results given in the table
to be strongly encouraging.   In this regard,
two points must be noted.  The first is that we must finally 
confront the problem of the normalization of the CFP.  Though we failed
to emphasize the point in our previous work, transition matrix elements 
calculated in CPC are 
independent of an overall rescaling of the normalization, provided it is the 
same for all states $|J\mu\nu\rangle$.  In our work we assumed unit 
normalization, as in the strong coupling core-particle model,
but this choice cannot be exact, as we know from our 
early work on spherical nuclei
\cite{K4}.  To make the appropriate corrections requires incorporating
into our algorithm a set of sum rules derived from the Fermion anti-commutation
relations.  This has not yet been done.  Second, to
achieve a result so close to the exact one, we must include 
(numerically significant) contributions from a large number of solutions
of the eigenvalue problem, including high-lying ones that play no role in the
fit to the known observables of the odd nuclei.  These points are relevant as 
well for the remainder of our discussion but will not be mentioned again.

4. {\it Self consistency: quadrupole matrix elements}.   
In the reference Hamiltonian
(\ref{ham}) there occur three quadrupole coupling constants.  In the case
under discussion, the experimental electric quadrupole matrix elements are 
in good
agreement with rigid rotor values.  For the CPC calculations we then 
assume  proportionality between
neutron and proton mass quadrupole elements, as expressed by the relation
\begin{equation} 
 \langle IMK |  Q^{\rm{n}} | I'M'K' \rangle = \eta ~\langle IMK | 
 Q^{\rm{p}} | I'M'K' \rangle  ,  \label{eta}
\end{equation}
where $\eta$ is a constant evaluated below. 
It follows that for the 
core-particle theory, we can work with the proton quadrupole moment alone
provided we introduce different effective coupling strengths for the neutron
and proton spectra, according to the equations
\begin{eqnarray}
   \kappa^{\rm{eff}}_{\rm{p}}
    &\equiv& \left( \kappa_{\rm{pp}} +\eta\, \kappa_{\rm{pn}} \right),
 \\
   \kappa^{\rm{eff}}_{\rm{n}}
    &\equiv& \left(\eta\, \kappa_{\rm{nn}} + \kappa_{\rm{pn}} \right) . 
\label{eq:eff_k}
\end{eqnarray}
Thus in fitting CPC to the data in the odd nuclei,
we are allowed to choose and indeed find slightly different values for the
effective coupling constants.  For these purposes and for the further
development, the actual value of $\eta$ is reflected only in the value that
has to be assigned to the effective coupling strengths.

\begin{figure}[h]
  \begin{center}
    \leavevmode
\centerline{\hbox{\psfig{figure=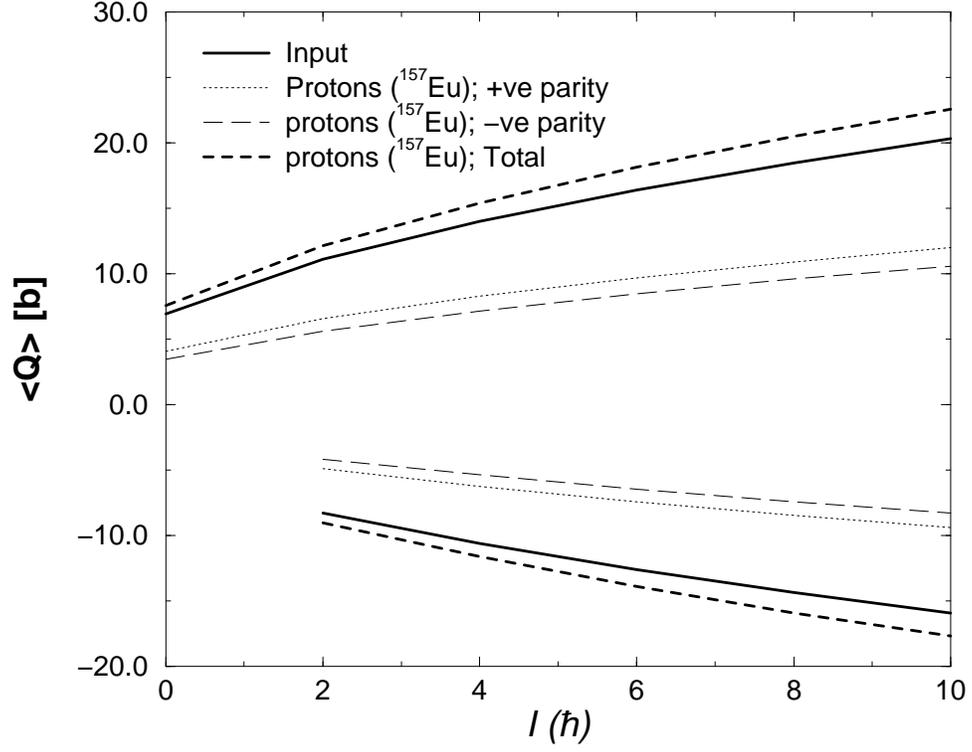,height=10cm}}    }
  \end{center}
  \caption{ Values of the matrix elements of $\hat{Q^{p}}$ for $^{158}$Gd.
The negative  values represent the diagonal elements 
$\langle IK=0 \| Q^{p} \| IK=0 \rangle $ whereas the positive values the 
off-diagonal elements \mbox{$\langle IK=0 \| Q^{p} \| I+2 \,K=0 \rangle $}. } 
  \label{fig:Q}
\end{figure}

Since we are dealing with operators of the form
\begin{equation}
\hat{Q}^i = \sum_{\alpha\gamma} q_{\alpha i,\gamma i} a^{\dag}_{\alpha i}
a_{\gamma i},   \label{cue}
\end{equation}
their core matrix elements are again quadratic sums in the same set of CFP
as enter the calculation of the number.   The first test of self-consistency
is that these sums have a ``shape'' consistent with the rigid rotor assumption,
as expressed in (\ref{eta}),
a test that is passed with flying colors.  This is seen partly 
from  Fig.~\ref{fig:Q}, which
emphasizes the fact that not only is the angular momentum dependence of the 
electric quadrupole matrix elements given correctly, but also their magnitudes
and signs.  From the fact that the neutron quadrupole matrix elements follow
parallel curves, we deduce the value $\eta=1.1$.

5. {\it Self consistency: pairing}.  For the pairing matrix elements, 
which are linear combinations of matrix elements of type
\begin{eqnarray}
&& \langle IMn(N-1)|a_{\bar{\alpha}}a_\alpha |I'M'n'(N+1)\rangle \nonumber\\
&& = \sum_{J\mu\nu} U^{\ast}_{J\mu\nu}(\alpha;IMn)V_{J\mu\nu}(\alpha;
I'M'n'),
\end{eqnarray}
there are again two tests.  First there is 
the requirement that the matrix elements be independent of angular momentum,
as was assumed in the input.   
Second we must reproduce the value of this matrix element.  For the neutron
pairing, the diagonal matrix elements vary monotonically between 1.859 
for $I=0$ and 1.814 for $I=8$, compared with the experimental value 
of 1.65 for the gap parameter.  The corresponding values for the proton
pairing are 1.672 for $I=0$ and 1.612 for $I=8$, compared with the experimental
value of 1.321.

6.  {\it Self consistency: moment of inertia}.    There remains the test of the
self consistency of the excitation spectrum.  This requires first 
that the diagonal
elements of (\ref{ham}) satisfy the rigid rotor equation
\begin{equation}
E_I = \langle IM|H|IM\rangle = \frac{I(I+1)}{2{\cal I}},
\end{equation}
and second that the experimental value of ${\cal I}$ be reproduced.
The first of these requirements is well-satisfied, and therefore
we can confine our attention to the moment of inertia.  In Table~\ref{tab:Moi},
we display not only the final calculated value of the moment
of inertia, but also the contributions of individual terms of $H$, broken
down according to nucleonic charge.  The absence of contributions from
the neutron quadrupole moment simply reflects the fact that we organized
the calculation so that the quadrupole term is expressed completely in terms 
the proton quadrupole operator and of the effective coupling strengths.
Notice that the major contribution comes overwhelmingly from the 
single-particle term and that the contribution of the quadrupole term 
is insignificant.  (This term then contributes only deformation energy.)
The self consistency is as close as one has a right to expect, in view
of the well-known effect of the quadrupole pairing interaction \cite{Hara},
which is not in our calculation.

7 {\it Final remarks}. The method of choosing parameters used
in fitting the odd-nucleus data has been thoroughly covered
in our previous work \cite{Pav1,Pav2,Pav3}.   In 
the present work we have checked a number of sum rules that
should be satisfied by a reasonably complete CPC calculation and found
agreement to within 10 percent.  These results imply that 
for a limited number of states we have demonstrated a new route
for passing directly from a spherical shell model to the properties of a
deformed rare-earth nucleus.  The immediate next steps are twofold:  to
add to the working Hamiltonian the additional simple interactions mentioned
in the text and to add to the algorithm a proper formulation for normalization
of the coefficients of fractional parentage.

\begin{table}[h]
  \begin{center}
  \begin{tabular}{ c | c | c }
    \hline 
     $\langle N \rangle$ & Actual & Calculated \\
     \hline 
     neutrons $+$ & 44  &  45.06 \\
     neutrons $-$ & 52  & 50.64 \\
     neutrons total& 94 & 95.70 \\   
     protons  $+$  & 38  &   36.26 \\
     protons $-$ & 26   &  28.90 \\
     protons total& 64  & 65.16 \\
     \hline 
    \end{tabular}
    \end{center}
\caption{ 
Particle number, actual and calculated, for $^{158}_{~\,64}$Gd.
}
\label{tab:Num} 
\end{table}

\begin{table}[t]
  \begin{center}
    \begin{tabular}{c|c|c|c|c|c}
      \hline 
     $E_2 - E_0$ & ${\cal I}_{\rm{sp}}$ & ${\cal I}_{\rm{Q}}$ &  ${\cal I}_{\rm{\Delta}}$ &  ${\cal I}_{\rm{total}}$ &  ${\cal I}_{\rm{exp}}$ \\
      \hline
      neutron $+$  & 0.00291  & - &  0.00121 &  & \\
      neutron $-$  &  0.00532 & -  &  -0.00236  & - &   \\
     neutron total & 0.00823  & -  &  -0.00157 & - &   \\
     proton $+$  &0.001943   &  0.00141 &   -0.00232 &  - &\\
     proton $-$    & 0.001272  & 0.000523  & 0.00194 & - & \\
     proton total     & 0.003215  & 0.001933  & -0.00126 & -  & \\ 
     Total         &    0.01145 &   -0.000643 &   0.001144 &  0.011951 & 0.0124 \\
     \hline 
     \end{tabular}
   \end{center}
\caption{Contributions to the moment of inertia arising from 
the different terms in the Hamiltonian. Here sp refers to the single particle 
contribution, $Q$ to the quadrupole contribution, and $\Delta$ to the pairing
term.}
\label{tab:Moi}
\end{table}

\end{document}